\documentstyle[prd,aps,floats,epsfig,times]{revtex}
\draft

\begin{document}
\twocolumn[\hsize\textwidth\columnwidth\hsize\csname@twocolumnfalse\endcsname
\title{Extended quintessence and the primordial helium abundance}
\author{Xuelei Chen$^{\star 1}$, 
Robert J. Scherrer$^{\dagger1,2}$, Gary Steigman$^{\ddag1,2}$}
\address{$^1$Department of Physics, The Ohio State University,
Columbus, OH~~43210}
\address{$^2$Department of Astronomy, The Ohio State University,
Columbus, OH~~43210}
\date{{\today}}
\maketitle

\begin{abstract}
In extended quintessence models, a scalar field which couples to
the curvature scalar $R$ provides most of the energy density of 
the universe.  We point out that such models can also lead naturally 
to a decrease in the primordial abundance of helium-4, relieving 
the tension which currently exists between the primordial helium-4 
abundance inferred from observations and the amount predicted by 
Standard Big Bang Nucleosynthesis (SBBN) corresponding to the 
observed deuterium abundance.  Using negative power-law potentials 
for the quintessence field, we determine the range of model parameters 
which can lead to an interesting reduction in the helium-4 abundance, 
and we show that it overlaps with the region allowed by other 
constraints on extended quintessence models.

\end{abstract}

 \vskip 2pc]

\section{Introduction}
A great deal of observational evidence currently points toward a
cosmological model with a nonzero cosmological constant $\Lambda$.   A
combination of the supernova Ia measurements \cite{SNIa}, measurements of
the baryon fraction in galaxy clusters \cite{WNEF}, and the location of
the first acoustic peak of the microwave background anisotropies
\cite{peak} suggests a model with $\Omega_{{\rm M}} = 0.3 - 0.4$ and
$\Omega_\Lambda = 0.6 - 0.7$.

A genuine cosmological constant is a disaster from the standpoint of
particle physics, in which the most natural value of $\Lambda$ is many
orders of magnitude larger than the ``observed'' value 
(see Ref.~\cite{Wein} for a review).
However, even if a plausible particle physics mechanism were developed
to produce such a small value of $\Lambda$, a second problem remains: 
why are $\Omega_\Lambda$ and $\Omega_{{\rm M}}$ comparable today? Since
$\Omega_\Lambda$ and $\Omega_{{\rm M}}$ scale very differently with the
cosmological expansion factor, this coincidence suggests that we live in
a very special epoch.  This problem has been dubbed the ``cosmological
constant coincidence problem'' to distinguish it from the more
fundamental cosmological constant problem \cite{ZWS,SWZ}.

A possible solution to the coincidence problem is to assume that the
apparent cosmological constant is not, in fact, a true constant vacuum
energy density, but instead is due to the energy density of a scalar
field $\phi$, a possibility which has come to be
known as {\em quintessence}. It might then be possible to couple the
behavior of this field to the background matter density in such way as
to achieve `naturally' the desired result, namely $\Omega_\phi \approx
\Omega_{{\rm M}}$ today.  Of course, in this case, the scalar field may
have an equation of state intermediate between matter and a cosmological
constant. This possibility was explored in some detail by Zlatev et al.
\cite{ZWS} (see also reference \cite{SWZ}) who argued that a certain
class of solutions (``tracking solutions'') will evolve toward the
desired behavior independent of the initial conditions.  Although such
models still require fine tuning to produce $\Omega_\phi \approx
\Omega_{{\rm M}}$ today \cite{LS},
they seem to be pointing in a more plausible
direction toward this result.

These investigations assumed minimally-coupled scalar fields, but 
it was soon realized that coupling the scalar field to the curvature 
scalar $R$ opens up another range of possibilities, dubbed ``extended 
quintessence.''  Uzan \cite{Uzan} examined the general case of a 
non-minimally coupled scalar field evolving in an exponential or 
power-law potential, and Amendola \cite{Amendola} explored a general 
class of couplings and potentials.  Non-minimally coupled models 
suffer from the potential problem that the gravitational constant 
$G$ varies in time \cite{Amendola,Chiba}, but they have nonetheless 
received a great deal of recent attention as possible models for 
quintessence \cite{pbm} - \cite{bmp}.

In this paper, we point out an interesting consequence of non-minimally 
coupled quintessence models:  under some circumstances these models 
can lead to a reduction in the primordial helium abundance.  This is 
of interest because of a ``tension'' between the recent estimate of 
the primordial deuterium abundance, $D/H = 3.0 \pm 0.4 \times 10^{-5}$ 
\cite{omeara}, which leads to an estimate for the baryon/photon ratio 
of $\eta = 5.6 \pm 0.5 \times 10^{-10}$, and the corresponding SBBN 
predicted primordial helium-4 abundance of $Y_P = 0.248 \pm 0.001$.
In contrast, the actual primordial helium abundance is likely lower.
For example, Olive and Steigman found \cite{Oliver-Steigman} 
\begin{equation}
Y_P = 0.234 \pm 0.003 ({\rm stat.}),
\end{equation}
while Izotov and Thuan obtained a higher value \cite{Izotov-Thuan}
\begin{equation}
Y_P = 0.244 \pm 0.002 ({\rm stat.}).
\end{equation}

While this apparent discrepancy is insufficient to discard SBBN, 
it certainly represents a ``tension'' in the model.  Furthermore, 
in addition to the three standard model neutrinos, a sterile neutrino 
may be needed to explain the results from neutrino oscillation 
experiments \cite{neutrino}. If either or both of these were confirmed, 
or if there are any other light particles in the Universe, the breach 
between theory and observation on $~^{4}\mathrm{He}$ could become 
even wider.

In light of these SBBN results, any ``natural'' mechanism which 
lowers the primordial helium-4 abundance must be regarded as 
interesting.  In this paper we show that extended quintessence 
models provide just such a mechanism.  In Ref. \cite{pbm}, Perrotta 
et al. investigated the BBN constraint on the extended quintessence 
model.  They showed that the model is not ruled out by increasing 
helium, but they did not consider the more interesting possibility 
of decreasing helium.

In the next section, we review the evolution of the scalar field
in these models and calculate the reduction in the abundance of 
helium-4 as a function of the model parameters.  We determine 
whether an interesting reduction is consistent with other 
constraints on extended quintessence. Our results are summarized 
in Sec. 3.  We shall use units in which $\hbar=c=8\pi G_{N}=1$ 
throughout, unless otherwise stated.

\section{The Non-Minimally Coupled Quintessence model}
The action for an extended quintessence model
is given by \cite{Uzan,pbm}
\begin{equation}
S=\int d^{4} x \sqrt{-g} \left[f(R,Q)-\frac{1}{2} \omega(Q) Q^{;\mu} 
Q_{;\mu} -V(Q) + L_{\rm fluid}\right],
\label{NMCeq}
\end{equation}
where $R$ is the curvature scalar of the space-time and $Q$ is a
scalar field.
We adopt the Non-Minimally Coupled (NMC) model of Refs.
\cite{pbm,bmp},
in which
$\omega(Q)=1$, $f(Q,R)=\frac{1}{2}F(Q) R$, and $F(Q)$ takes the form,
\begin{equation}
\label{FQ}
F(Q)=1-\xi(Q^2 - Q_0^{2}),\quad
\end{equation}
where $Q_0$ is the value of $Q$ today.
(The zero subscript will refer throughout
to quantities evaluated at the present).
Eq. (\ref{FQ}) ensures
that $F(Q) = 1$ today.  

The coupling constant $\xi$ in the NMC quintessence model is
constrained by solar system limits on $\omega_{JBD}$,
the Jordan-Brans-Dicke parameter \cite{pbm}:
\begin{equation}
\omega_{JBD} = \frac{F_0}{(\partial F / \partial Q)_0^2} >500,
\label{xiEq}
\end{equation}
as well as the experimental limit on variation of the gravitational constant
\cite{pbm}:
\begin{equation}
|\dot G/G|_0 = |\dot F/F|_0<10^{-11} {\rm yr}^{-1}.
\end{equation}
Using the form for $F(Q)$ from Eq. (\ref{FQ}), these limits
translate into:
\begin{equation}
\label{solarsystem}
\xi Q_0 < 0.022,
\end{equation}
from the solar system constraint,
and
\begin{equation}
\label{varyG}
2 \xi Q_0 \dot Q_0 < 10^{-11} {\rm yr}^{-1},
\end{equation}
from the limits on the time variation of $G$.

In this model, the expansion rate of the Universe is
\begin{equation}
\label{Hequation}
H^2 = \frac{1}{3F} 
\left(\rho_f + \frac{1}{2} \dot{Q}^2 +V(Q) + 6 H \xi Q \dot Q
\right),
\label{Heq}
\end{equation}
where $\rho_f = \rho_m +\rho_r$ is the density contribution from 
matter and radiation (including neutrinos).
The evolution of $Q$ is governed by
\begin{equation}
\label{qevol}
\ddot{Q}+3H\dot{Q} + {\partial V \over \partial Q} + \xi R Q = 0.
\label{phieq}
\end{equation}
The scalar curvature
is given by
\begin{eqnarray}
R&=&6(\dot{H}+2H^2 )\nonumber\\
&=&\frac{1}{F}\left(\rho_f -3 p_f - \dot{Q}^2 + 
4 V + 6\xi(\dot Q^2 + Q \ddot Q + 3H Q \dot Q )\right).  
\label{Req}
\end{eqnarray}

As a specific case, we will consider the inverse power law
potential for $Q$:
\begin{equation}
V(Q)=V_{0} Q^{-\alpha}.
\end{equation}
However, our final results can be generalized to other forms for
the potential.  It is known that with this type of potential, the 
quintessence energy redshifts more slowly than the radiation in 
the radiation-dominated era, and more slowly than the matter in 
the matter-dominated era, ultimately dominating at late times 
\cite{ZWS,SWZ,LS,bmp,ratra}, thus providing an explanation for 
the observed accelerating expansion of the Universe \cite{SNIa}.
The vacuum energy observed today is determined by $V_0$ but is 
not (or is only weakly) dependent on the initial value of $Q$, 
thus, partly addressing the ``fine tuning problem'' \cite{ZWS,SWZ}.

\begin{figure}[htbp]
\begin{center}
\epsfig{file=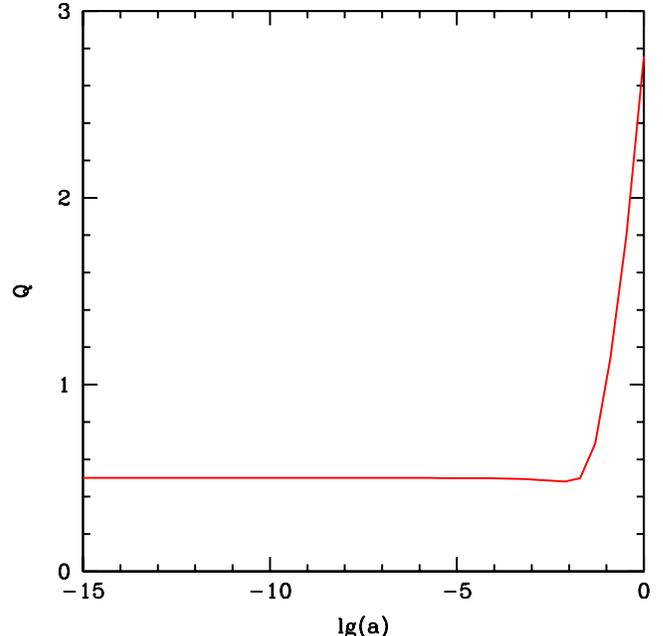,width=3.5in}
\caption{The evolution of $Q$ for a potential $V = V_0 Q^{-4}$,
where $V_0$ is chosen to satisfy the boundary conditions given
in equations (\ref{omegacond}), and $\xi = 0.007$.}
\label{Qevol}
\end{center}
\end{figure}

\begin{figure}[htbp]
\begin{center}
\epsfig{file=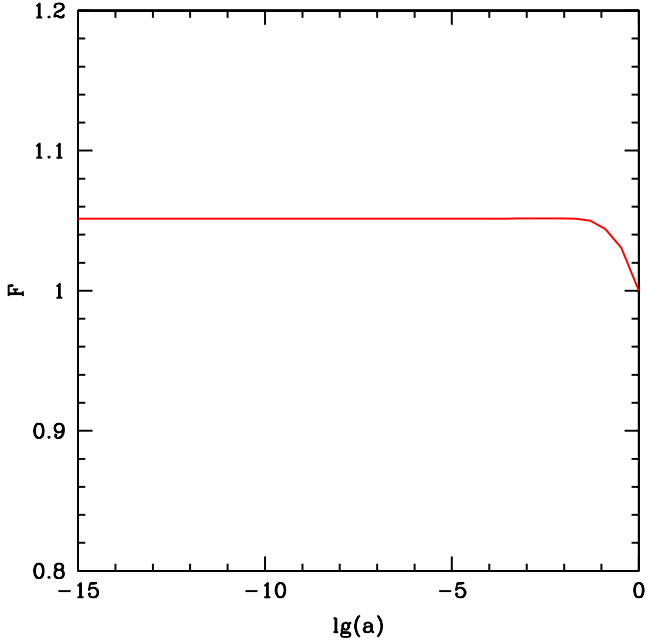,width=3.5in}
\caption{As in Fig. \ref{Qevol}, for the evolution of $F$.}
\label{Fevol}
\end{center}
\end{figure}

A viable NMC cosmological model must satisfy the following boundary
conditions at $a=1$:
\begin{itemize} 
\item The effective gravitational constant today must be equal 
to the measured value, which translates to the requirement 
\begin{equation}
F|_{a=1} =1.
\label{Fcond}
\end{equation} 
This condition is satisfied automatically by our definition in 
equation (\ref{FQ}).
\item After evolving the model from an early epoch to the present 
one, we must recover the present value of cosmological parameters,
which we take to be:
\begin{equation}
\label{omegacond}
\Omega_{M0} = 0.3, \qquad H_0 = 65~{\rm km/sec/Mpc}
\label{Hcond} 
\end{equation}
\end{itemize}
where, as noted above, we take $\Omega_0=1$.
As an example, we show the evolution of $Q$, $F$, and the densities
of interest which satisfy our boundary conditions in Figs. 1-3,
for the case $V = V_0 Q^{-4}$, $\xi=0.007$.

The general behavior of the NMC quintessence model with this potential
was discussed
in Ref.~\cite{bmp}; here we review it briefly.  At sufficiently
early times, the $\xi R Q$ term in equation (\ref{qevol})
dominates the $\partial V /\partial Q$ term, and
the field $Q$ settles down to a slow 
roll regime, with an effective potential 
$\xi R Q$ balanced by 
a ``frictional force'' $-3H\dot{Q}$.
(This dominance of the $\xi R Q$ term has been dubbed
the ``R-boost" \cite{bmp}).
In this regime, $Q$ is nearly constant (as seen in Fig.~\ref{Qevol}), and
the potential term $V$ is sub-dominant; $V$  
affects neither the expansion nor the $Q$-evolution significantly.
The slow roll holds until $V$ 
becomes significant, after which $Q$ begins to roll fast, and 
the quintessence kinetic and potential energy come to dominate.
In this regime, the $\xi R Q$ term can be neglected, and
the field behaves as a minimally coupled ``tracker'' field
with a negative power law potential \cite{ZWS,SWZ,LS,bmp,ratra}.
The results we show in Figs. 1-3 are in agreement with those
of Ref. \cite{bmp}.

Depending on the choice of the initial conditions, the history of the 
$Q$ field may have a few more twists than described above; we 
refer the reader to Ref.~\cite{bmp} for more details. For our purpose,
it is sufficient to note that the early slow roll and the late 
tracker behavior are common features
for most of these models, and we will confine
our investigation to the range of parameters for which
this behavior holds.

\begin{figure}[tbp]
\begin{center}
\epsfig{file=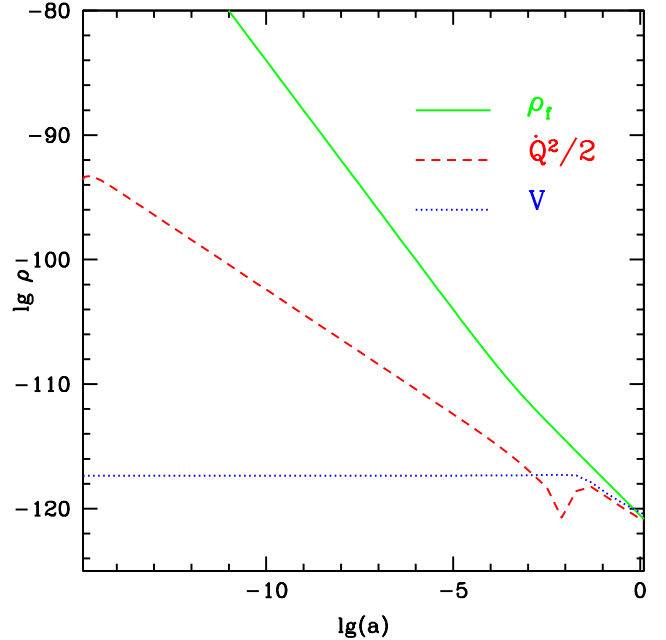,width=3.5in}
\caption{As in Fig.~\ref{Qevol}, 
for the evolution of the fluid and Q-field energy densities.}
\label{rhoevol}
\end{center}
\end{figure}

\begin{figure}[htbp]
\begin{center}
\epsfig{file=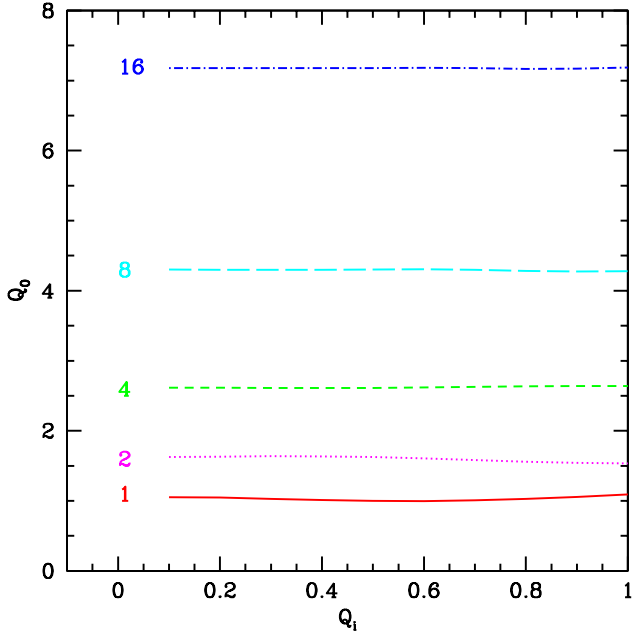,width=3.5in}
\caption{$Q_0$ as a function of $Q_i$ for
various values of $\alpha$, shown on the left side
of each curve, for $\xi=0.02$.}
\label{QiQ0alpha}
\end{center}
\end{figure}

Our model appears to have four free parameters:  $V_0$, $\alpha$,
$\xi$, and $Q_i$ (the initial value of $Q$), which together determine $Q_0$.
Note, however, that $Q_0$ is almost completely independent of $Q_i$,
as a consequence of the tracker behavior of the model \cite{ZWS,SWZ}.
This is displayed in Fig. \ref{QiQ0alpha} 

\begin{figure}[hbp]
\begin{center}
\epsfig{file=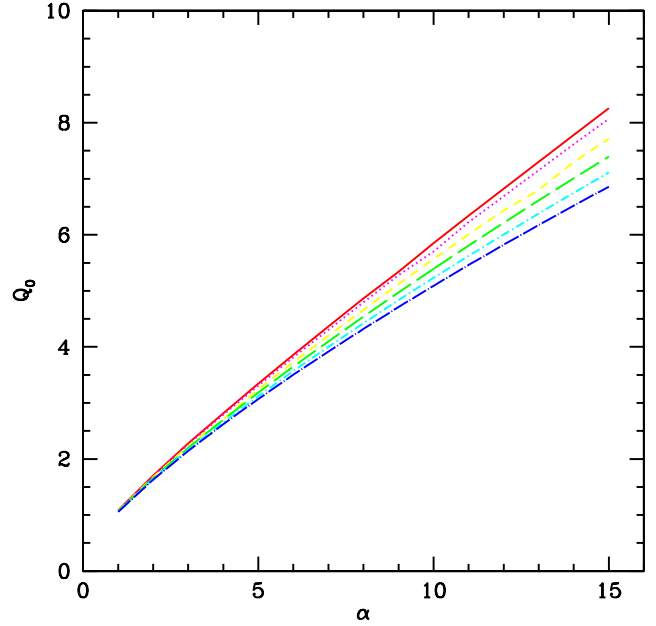,width=3.5in}
\caption{$Q_0$ as a function of $\alpha$, for (from top to
bottom) $\xi= 0.002, 0.004, 0.008, 0.012, 0.016, 0.02$.}
\label{alpha-Q0}
\end{center}
\end{figure}

Hence, $Q_0$ is effectively only a function of $\alpha$, $V_0$,
and $\xi$.  However, once we fix $\alpha$, and $\xi$, the value
of $V_0$ is fixed by boundary
conditions of Eq. (\ref{omegacond}).  So effectively,
$\alpha$ and $\xi$, along with the boundary conditions, fix
$Q_0$.  This dependence is shown in Fig. \ref{alpha-Q0}.  It is
obvious from this figure that $Q_0$ is also nearly
independent of $\xi$.  This follows from the fact that
the term containing $\xi$ in Eq. (\ref{qevol}) is subdominant
once tracking behavior starts.  Hence, it is a good approximation
to take $Q_0$ to be a function of $\alpha$ alone for boundary
conditions fixed as in Eq. (\ref{omegacond}).

\begin{figure}[tbp]
\begin{center}
\epsfig{file=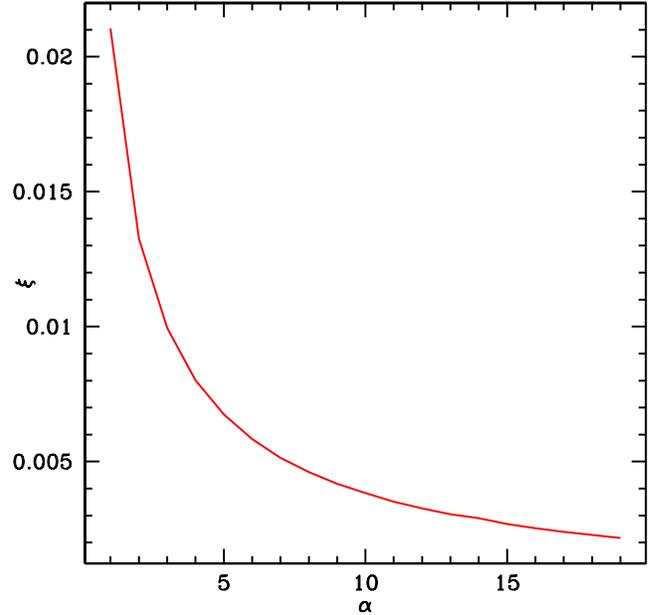,width=3.5in}
\caption{The strongest coupling for a given $\alpha$ allowed 
by solar system experiments.}
\label{alpha-xi}
\end{center}
\end{figure}

Since
the solar system limit (Eq. \ref{solarsystem}) is a constraint
on $\xi$ and $Q_0$, it translates into a constraint on $\xi$
for a given value of $\alpha$.  This limit is shown in Fig.
\ref{alpha-xi}.  For the models we consider here, the limit
from equation (\ref{varyG}) is subdominant and can be ignored.

Now consider the effect on the expansion rate.  Let
$H$ be the Hubble parameter in our NMC quintessence model,
and let $\bar H$ be 
the Hubble parameter in a fiducial (no quintessence) model,
which we take to be a standard $\Lambda$CDM model with
$\Omega_\Lambda = 0.7$, $\Omega_0 = 1$.
(Our specific choice of fiducial model is relevant only at late times
and does not affect our BBN calculations, but we choose
this particular model for definiteness).
In 
order to reduce the helium abundance, we would like $H<\bar{H}$, 
which implies $F>1$, during BBN. However, when $V$ dominates,  
$Q$ rolls ``down hill'' in the positive direction (assume $Q>0$), 
so $Q_0 > Q_{BBN}$. Therefore, we find $F>1$ if and only if $\xi 
>0$.

\begin{figure}[tb]
\begin{center}
\epsfig{file=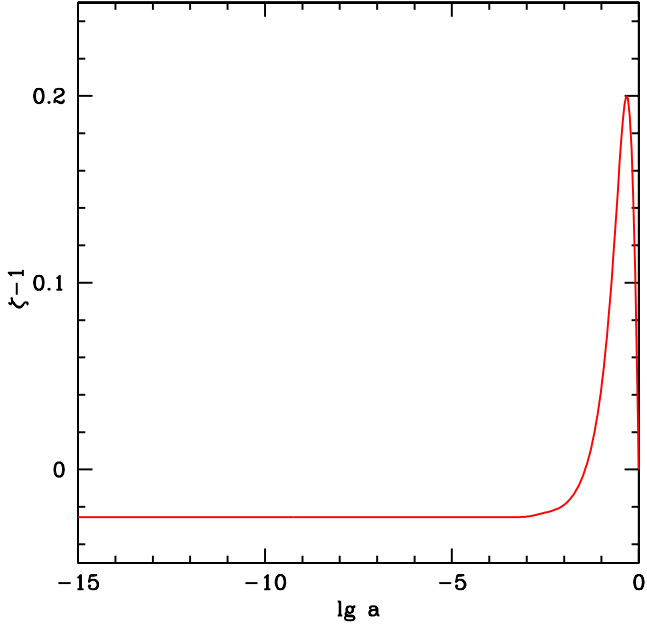,width=3.5in}
\caption{As in Fig. 1, for the evolution of the speed-up factor $\zeta$.}
\label{zetaevol}
\end{center}
\end{figure}

The change in expansion rate can be parametrized by a speed-up 
factor $\zeta$:
\begin{equation}
\zeta \equiv \frac{H}{\bar{H}}.
\end{equation}
The evolution of $\zeta-1$ in the model of Figs. 1-3 is
shown in Fig.~\ref{zetaevol}.  The region at the right side
of the graph for which $\zeta - 1 > 0 $ arises after the quintessence
field enters the tracker regime; in this regime the
scaling of $\rho_Q$ with $a$ is different than the scaling
of  $\rho_\Lambda$ in our fiducial model.  This behavior is, however, irrelevant
for BBN; the value of $\zeta-1$ during BBN
depends only on the radiation content of our fiducial model.

\begin{figure}[tbp]
\begin{center}
\epsfig{file=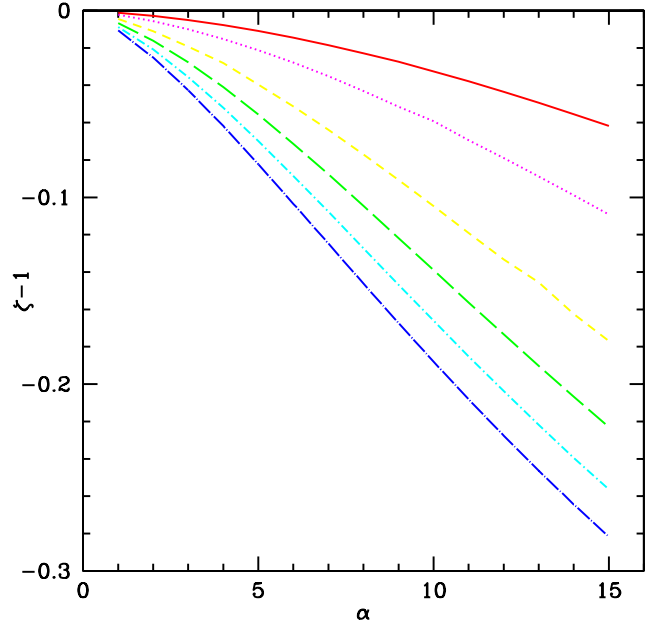,width=3.5in}
\caption{The value of $\zeta-1$ during nucleosynthesis as a
function of $\alpha$; the five 
curves are, from top to bottom, $\xi= 0.002, 0.004, 0.008, 
0.012, 0.016, 0.02$.}
\label{alphazeta-xif}
\end{center}
\end{figure}

During BBN, $\rho_f$ dominates the density in equation (\ref{Hequation}),
so the speed-up factor is determined primarily by the effective 
gravitational constant $F$.
Since the $Q$ field is almost frozen during the radiation
dominated era, we have $Q \approx Q_i$, so
\begin{equation}
\zeta - 1 \approx -{1 \over 2} (F-1) \approx {1\over 2}\xi(Q_i^2 - Q_0^2),
\end{equation}
for $\zeta - 1 \ll 1$.
We are primarily interested in the case $Q_i << Q_0$, because
it is in this regime one expect a large negative $\zeta-1$.
In this case, we get
\begin{equation}
\label{zeta}
\zeta-1 \approx -{1\over 2}\xi Q_0^2.
\end{equation}
As we have seen, $Q_0$ is effectively only a function of $\alpha$,
with a slight residual dependence on $\xi$ for the boundary
conditions fixed in Eq. (\ref{omegacond}).  Hence, the speed-up factor
will be a function of $\xi$ and $Q_0$ alone; we expect
$\zeta-1 \propto \xi$ from Eq. (\ref{zeta}), and Fig. \ref{alpha-Q0}
shows that $Q_0$, and hence the magnitude of $\zeta-1$
should increase with increasing $\alpha$.  The dependence
of $\zeta - 1$ as a function of $\xi$ and $\alpha$
is shown in Fig.~\ref{alphazeta-xif}. 


Now consider the effect that the change in $H$ will have on 
the primordial helium abundance.  For all values of $\alpha$ 
considered here ($\alpha < 15$), $F$ is essentially constant 
during BBN, so that we can take $\zeta - 1$ to be constant.  
It is then straightforward to calculate the change in the 
predicted primordial helium abundance (compared to SBBN).
For small changes in the expansion rate at BBN virtually
all the neutrons available when BBN begins are incorporated
in helium-4, so the helium abundance is directly related to
the neutron abundance.  The faster the expansion the more
neutrons are available and the more helium synthesized.  
A slower expansion has the opposite effect.  For small 
deviations from the SBBN expansion rate

\begin{equation}
\Delta Y \approx 0.08 (\zeta^2 -1) \approx 0.16 (\zeta -1 )
\approx - 0.08 \xi Q_0^2.
\end{equation}

\begin{figure}[tbp]
\begin{center}
\epsfig{file=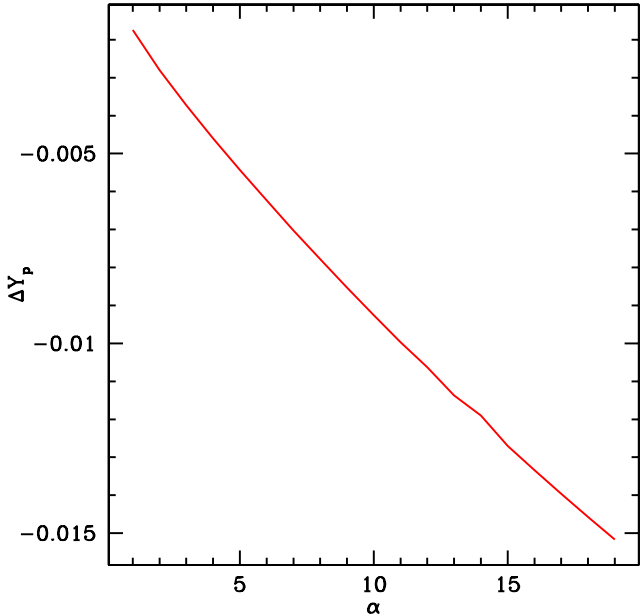,width=3.5in}
\caption{The change in the primordial helium,
$\Delta Y$, versus $\alpha$, with strongest allowed
coupling $\xi$.}
\label{alphaYp}
\end{center}
\end{figure}

In general, $\Delta Y$ is a function of both $\alpha$ and
$\xi$.  However, we would like to find the largest possible
value of $\Delta Y$ that is consistent with the solar system
bounds of Eq. (\ref{solarsystem}).  Since $\Delta Y$ increases
with $\xi$, we use the largest possible $\xi$ for a given
$\alpha$ (shown in Fig. \ref{alpha-xi}) to determine $\Delta Y$
for a given $\alpha$.  This is shown in Fig. \ref{alphaYp}.

It is obvious from this figure that $- \Delta Y$ is an increasing
function of $\alpha$.  For instance, with $\alpha = 10$, we find 
$\Delta Y \approx - 0.01$, corresponding to $\xi = 0.004$.  Often 
the speed up in the expansion rate is parameterized in terms of 
an equivalent number of ``extra" neutrinos, $\Delta N_{\nu}$.

\begin{equation}
\Delta N_{\nu} \equiv {43\over 7}(\zeta^{2} - 1) \approx
-{43\over 7} \xi Q_0^2.
\end{equation}
This reduction in $Y_P$ from its SBBN value corresponds to a 
reduction of $N_{\nu}$ from its standard model value of 3 by
$\Delta N_{\nu} \approx 0.74$.  There is another, subdominant
effect of a slower expansion in that there will be more time
available to destroy deuterium as well as to synthesize
$^{7}$Be which will later electron capture and add
to the abundance of $^{7}$Li.  As a result, the same deuterium
and lithium abundances will correspond to a slightly smaller
baryon-to-photon ratio which, in turn, will yield a slightly
smaller predicted helium abundance.
 
\section{Discussion and Conclusion}
We have investigated the effect of NMC quintessence model on BBN. 
In models with $\xi>0$, the expansion rate is smaller during BBN, 
and less helium is produced. The amount of this reduction depends 
primarily on the coupling constant and the slope of the potential, 
and also very weakly on the initial value of the field.  However, 
the coupling constant $\xi$ is limited by solar system experiments.  
We find that for negative power-law potentials of the form $V = V_0 
Q^{-\alpha}$, the magnitude of the reduction increases with $\alpha$.  
This is not a desirable state of affairs from the point of view of 
reducing primordial helium.  The reason is that negative power-law 
potentials give an equation of state $w = p/\rho$ in which $w$ 
approaches 0 as $\alpha$ increases.  In Figure 10 is shown the 
relation between $w$ and $\alpha$.  For instance, $\alpha = 6$ 
corresponds to $w = -0.43$, and a variety of observations argue 
against a value of $w$ much larger than this (see, e.g., Ref. 
\cite{wcos}).  If we take $\alpha \le 6$ as a reasonable upper 
bound on such models, then we find $\Delta Y \le 0.0061$.  This 
is an interesting reduction in primordial helium (from $Y_P = 
0.248$ to $Y_P = 0.242$), albeit it requires us to push all of 
the parameters in the model to the extreme acceptable limits.  
For example, while a reduction as small as $\Delta Y \approx 
0.002 - 0.004$ would be sufficient to reconcile the O'Meara et al.
deuterium abundance \cite{omeara} with the Izotov and Thuan 
helium value \cite{Izotov-Thuan}, a much larger reduction would 
be required if the Olive and Steigman \cite{Oliver-Steigman} 
helium abundance were adopted.

If there is indeed a breach between the observed helium abundance 
and the predictions of SBBN it could be healed by this NMC model. 
Alternatively, a reduction in $Y_P$ of 0.0061 corresponds to a shift 
in the equivalent number of neutrinos by $\Delta N_{\nu} = - 0.47$.

\begin{figure}[tbp]
\begin{center}
\epsfig{file=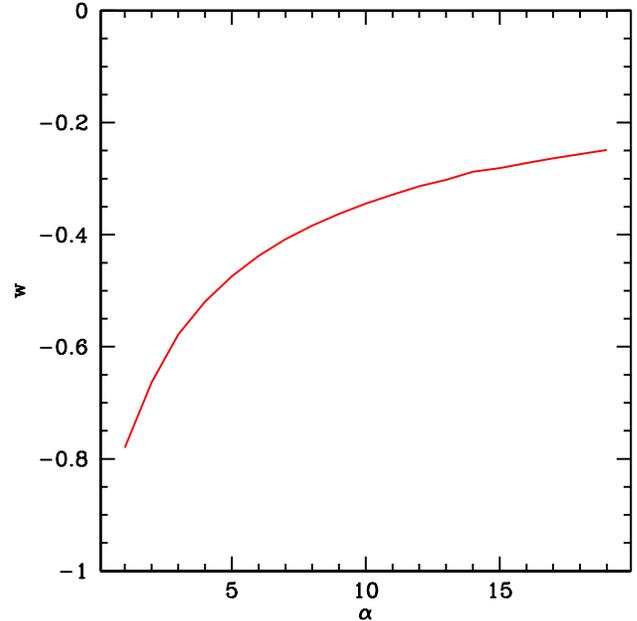,width=3.5in}
\caption{$w$ versus $\alpha$, with strongest allowed
coupling $\xi$.}
\label{alphaw}
\end{center}
\end{figure}

We have shown that certain classes of NMC quintessence models
lead to a ``natural" reduction in the primordial helium abundance.
Although we have concentrated on the negative power law potentials,
our results generalize easily to other potentials.  In particular,
one of the major problems with exponential potentials is that
they lead to an overproduction of helium \cite{expQmodel}.  This
problem could be ameliorated in the way we have suggested.
Of course, the tracker solution
for such models generically predicts $w = 0$ at present \cite{expQmodel}
and thus these models have problems matching other observations.  We also
expect that these results would be changed if we used
a different coupling between $Q$ and $R$ than that in Eq. (\ref{FQ}).
Thus, it might be possible to obtain even larger
values for $\Delta Y$ in such models.

\acknowledgments
X.C., R.J.S., and G. S. are supported by the DOE (DE-FG02-91ER40690).

\newcommand\AJ[3]{~Astron. J.{\bf ~#1}, #2~(#3)}
\newcommand\APJ[3]{~Astrophys. J.{\bf ~#1}, #2~ (#3)}
\newcommand\apjl[3]{~Astrophys. J. Lett. {\bf ~#1}, L#2~(#3)}
\newcommand\ass[3]{~Astrophys. Space Sci.{\bf ~#1}, #2~(#3)}
\newcommand\cqg[3]{~Class. Quant. Grav.{\bf ~#1}, #2~(#3)}
\newcommand\mnras[3]{~Mon. Not. R. Astron. Soc.{\bf ~#1}, #2~(#3)}
\newcommand\mpla[3]{~Mod. Phys. Lett. A{\bf ~#1}, #2~(#3)}
\newcommand\npb[3]{~Nucl. Phys. B{\bf ~#1}, #2~(#3)}
\newcommand\plb[3]{~Phys. Lett. B{\bf ~#1}, #2~(#3)}
\newcommand\pr[3]{~Phys. Rev.{\bf ~#1}, #2~(#3)}
\newcommand\PRL[3]{~Phys. Rev. Lett.{\bf ~#1}, #2~(#3)}
\newcommand\PRD[3]{~Phys. Rev. D{\bf ~#1}, #2~(#3)}
\newcommand\prog[3]{~Prog. Theor. Phys. {\bf ~#1}, #2~(#3)}
\newcommand\MeV{\mathrm{MeV}}


\begin{references}
\bibitem[\star]{byline} Electronic address: xuelei@pacific.mps.ohio-state.edu.
\bibitem[\dagger]{byline} Electronic address: scherrer@pacific.mps.ohio-state.edu.
\bibitem[\ddag]{byline} Electronic address: steigman@pacific.mps.ohio-state.edu.

\bibitem{SNIa} A. G. Riess et al., \AJ{116}{1109}{1998}; P. M. Garnavich et
al., \APJ{509}{74}{1998};
S. Perlmutter et al., 
\APJ{517}{565}{1999}.
\bibitem{WNEF} S. D. M. White, J. F. Navarro, A. E. Evrard and
	C. S. Frenk, Nature {\bf 366}, 429 (1993).
\bibitem{peak} P. de Bernardis et al., Nature {\bf 404}, 955 (2000);
S. Hanany et al., Astrophys. J. Lett. {\bf 545}, 5 (2000);
A. H. Jaffe, astro-ph/0007333.
\bibitem{Wein} S. Weinberg, Rev. Mod. Phys. {\bf 61}, 1 (1989).
\bibitem{ZWS} I. Zlatev, L. Wang and P. J. Steinhardt, \prl {\bf 82},
	896 (1999). 
\bibitem{SWZ} P. J. Steinhardt, L. Wang, and I. Zlatev, \prd {\bf 59},
        123504 (1999).
\bibitem{LS} A. R. Liddle and R.J. Scherrer, \prd {\bf 59}, 023509 (1999).
\bibitem{Uzan} J.-P. Uzan, \prd {\bf 59}, 123510 (1999).
\bibitem{Amendola} L. Amendola, \prd {\bf 60}, 043501 (1999).
\bibitem{Chiba} T. Chiba, \prd {\bf 60}, 083508 (1999).
\bibitem{pbm} F. Perrotta, C. Baccigalupi, S. Matarrese, \prd {\bf 61},
023507 (2000).
\bibitem{hw} D.J. Holden and D. Wands, \prd {\bf 61}, 043506 (2000).
\bibitem{bmp} C. Baccigalupi, S. Matarrese, F. Perrotta, \prd {\bf 62},
123510 (2000).

\bibitem{omeara} J.M. O'Meara, D. Tytler, D. Kirkman, N. Suzuki,
J.X. Prochaska, A.M. Wolfe, Astrophys. J., submitted, astro-ph/0011179.

\bibitem{Oliver-Steigman} K. A. Olive and G. Steigman,
 Astrophys. J. Supp. {\bf 97}, {49} (1995).

\bibitem{Izotov-Thuan} Y. I. Izotov and T. X. Thuan, 
\APJ{500}{188}{1998}.

\bibitem{neutrino} For a review, see e.g. P. Langacker,
Nucl. Phys. B(Proc. Suppl.){\bf 77}, 241(1999).  

\bibitem{ratra}
B. Ratra and P. J. E. Peebles, \PRD{37}{3406}{1988}

\bibitem{wcos} L. Wang, R.R. Caldwell, J.P. Ostriker, and P.J.
Steinhardt, Astrophys. J. {\bf 530}, 17 (2000).

\bibitem{PDG} D. E. Groom et al. (PDG), Euro. Phys. J. {\bf C15}, 1 (2000).

\bibitem{expQmodel} P. G. Ferreira and M. Joyce, \PRL{79}{4740}{1997},
E. J. Copeland, A. Liddle and D. Wands, \PRD{57}{4686}{1998}.

\end{references}
\end{document}